\PassOptionsToPackage{unicode}{hyperref}
\PassOptionsToPackage{hyphens}{url}
\documentclass[
]{article}
\usepackage{xcolor}
\usepackage{amsmath,amssymb}
\usepackage{iftex}
\ifPDFTeX
  \usepackage[T1]{fontenc}
  \usepackage[utf8]{inputenc}
  \usepackage{textcomp} % provide euro and other symbols
\else % if luatex or xetex
  \usepackage{unicode-math} % this also loads fontspec
  \defaultfontfeatures{Scale=MatchLowercase}
  \defaultfontfeatures[\rmfamily]{Ligatures=TeX,Scale=1}
\fi
\usepackage{lmodern}
\ifPDFTeX\else
\fi
\IfFileExists{upquote.sty}{\usepackage{upquote}}{}
\IfFileExists{microtype.sty}{% use microtype if available
  \usepackage[]{microtype}
  \UseMicrotypeSet[protrusion]{basicmath} % disable protrusion for tt fonts
}{}
\makeatletter
\@ifundefined{KOMAClassName}{% if non-KOMA class
  \IfFileExists{parskip.sty}{%
    \usepackage{parskip}
  }{% else
    \setlength{\parindent}{0pt}
    \setlength{\parskip}{6pt plus 2pt minus 1pt}}
}{% if KOMA class
  \KOMAoptions{parskip=half}}
\makeatother
\usepackage{longtable,booktabs,array}
\usepackage{calc} % for calculating minipage widths
\usepackage{etoolbox}
\makeatletter
\patchcmd\longtable{\par}{\if@noskipsec\mbox{}\fi\par}{}{}
\makeatother
\IfFileExists{footnotehyper.sty}{\usepackage{footnotehyper}}{\usepackage{footnote}}
\makesavenoteenv{longtable}
\usepackage{graphicx}
\makeatletter
\newsavebox\pandoc@box
\newcommand*\pandocbounded[1]{% scales image to fit in text height/width
  \sbox\pandoc@box{#1}%
  \Gscale@div\@tempa{\textheight}{\dimexpr\ht\pandoc@box+\dp\pandoc@box\relax}%
  \Gscale@div\@tempb{\linewidth}{\wd\pandoc@box}%
  \ifdim\@tempb\p@<\@tempa\p@\let\@tempa\@tempb\fi% select the smaller of both
  \ifdim\@tempa\p@<\p@\scalebox{\@tempa}{\usebox\pandoc@box}%
  \else\usebox{\pandoc@box}%
  \fi%
}
\def\fps@figure{htbp}
\makeatother
\ifLuaTeX
  \usepackage{luacolor}
  \usepackage[soul]{lua-ul}
\else
  \usepackage{soul}
\fi
\usepackage{bookmark}
\IfFileExists{xurl.sty}{\usepackage{xurl}}{} % add URL line breaks if available
\hypersetup{
  hidelinks,
  pdfcreator={LaTeX via pandoc}}

\author{}
\date{}

\begin{document}

\textbf{IyàwóBench: A Benchmark for Evaluating Large Language Model
Clinical Triage Accuracy on Undifferentiated Febrile Illness in Nigerian
Primary Health Settings}

Anthonio Oladimeji Gabriel\textsuperscript{*}, Dimeji Abdulsobur
Olawuyi, Oloruntoba Ajayi, Temiloluwa Aderemi

\emph{*} \textbf{Centre for Clinical Intelligence and Safety, Iyawo}

Correspondence: anthoniooladimeji11@gmail.com

\textbf{Abstract}

\textbf{Background.} Undifferentiated febrile illness is the leading
cause of primary care outpatient visits in Nigeria, yet no validated
benchmark exists for evaluating large language model (LLM) clinical
triage reasoning in West African primary health settings.

\textbf{Methods.} We introduce IyàwóBench v1.0, a dataset of 200
synthetic clinical vignettes across eight febrile illness categories
derived from statistical distributions of 1,200 real patient encounters
at 19 primary health centres (PHCs) in Oyo State, Nigeria. Six LLMs were
evaluated on structured triage classification across two metrics: triage
accuracy and safety score.

\textbf{Results.} All six models achieved 100\% safety scores (95\% CI:
96.4-100.0\%), never downgrading a critical REFER NOW case to TREAT
HERE. Triage accuracy varied substantially: Claude Sonnet
(claude-sonnet-4-5) 67.5\% (95\% CI: 60.8-73.7\%), Llama 4 Scout 59.5\%
(52.5-66.2\%), Llama 3.3 70B 43.0\% (36.2-50.0\%), and Llama 3.1 8B
39.0\% (32.4-45.9\%). Two models demonstrated near-zero accuracy
attributable to structured output non-compliance.

\textbf{Conclusions.} Modern LLMs exhibit safe triage behaviour but vary
substantially in structured clinical accuracy. Clinically engineered
systems with embedded WHO guidelines outperform general-purpose models
by up to 28.5 percentage points. IyàwóBench provides the first
reproducible evaluation framework for LLM clinical decision support in
West African primary care.


\textbf{1. Introduction}

Nigeria operates approximately 30,000 primary health centres (PHCs),
which constitute the first point of contact for the majority of the
country\textquotesingle s 220 million people with the formal health
system {[}1{]}. These facilities are staffed predominantly by Community
Health Extension Workers (CHEWs), with a national ratio of fewer than
one CHEW per 5,000 patients {[}2{]}. Undifferentiated febrile illness,
in which fever is the predominant symptom but the underlying aetiology
is not immediately apparent, accounts for an estimated 20 to 30 percent
of all PHC outpatient consultations {[}3{]}. The diagnostic challenge is
substantial: malaria, typhoid fever, bacterial meningitis, sepsis, and
pneumonia share overlapping early presentations, yet each requires
fundamentally different management. Bacterial meningitis and severe
sepsis carry case fatality rates exceeding 20 percent without prompt
hospital-level treatment {[}4,5{]}, while uncomplicated malaria resolves
with outpatient artemisinin-based combination therapy {[}6{]}.

Community health workers making triage decisions without clinical
decision support tools have been shown to misclassify febrile
presentations at rates that result in both dangerous under-referral and
unnecessary over-referral {[}7,8{]}. Algorithm-based clinical decision
support for Integrated Management of Childhood Illness (IMCI) in Burkina
Faso demonstrated 83 to 85 percent agreement with physician examination
findings, establishing proof of concept for electronic decision support
at the community health worker level {[}9{]}.

Large language models (LLMs) have attracted substantial interest as a
mechanism for delivering adaptive clinical decision support at scale in
low-resource settings {[}10,11{]}. Singhal et al. demonstrated that
large-scale language models can achieve expert-level performance on
United States medical licensing examination questions {[}12{]}. However,
performance on standardised medical examinations does not translate
directly to clinical triage accuracy in low- and middle-income country
(LMIC) settings, where disease prevalence, presentation patterns,
infrastructure constraints, and evidence-based guidelines differ
substantially from high-income country contexts {[}13{]}. Reported
deployments of LLM-powered clinical tools in sub-Saharan Africa have
encountered significant challenges: Babylon Health\textquotesingle s
Rwanda deployment was discontinued in 2023 following financial collapse
{[}14{]}, and a published evaluation of an LLM-powered clinical decision
support system across 16 primary care facilities in Kenya found that 7.8
percent of AI-generated management recommendations were actively harmful
{[}15{]}.

A fundamental barrier to evidence-based LLM deployment in LMIC primary
care is the absence of validated evaluation benchmarks grounded in local
disease burden, clinical infrastructure, and regional guidelines.
Existing benchmarks including MedQA {[}16{]}, MedMCQA {[}17{]}, and
PubMedQA {[}18{]} are derived predominantly from United States or Indian
medical licensing examinations and biomedical literature, and do not
reflect the diagnostic challenges encountered by a CHEW triaging a
febrile patient at a Nigerian PHC with a basic Android smartphone and
intermittent internet connectivity.

We introduce IyàwóBench, the first benchmark dataset designed to
evaluate LLM triage accuracy on undifferentiated febrile illness in
Nigerian primary health settings. IyàwóBench v1.0 comprises 200
synthetic clinical vignettes whose statistical distributions are derived
from 1,200 real patient encounters collected during the initial
deployment of Iyawo, an AI-powered clinical decision support platform,
across 19 PHCs in Oyo State, Nigeria, under formal approval from the Oyo
State Primary Healthcare Board. We evaluate six frontier and open-source
LLMs across two clinically meaningful metrics: triage accuracy and a
safety score measuring avoidance of dangerous clinical downgrades.

\textbf{2. Methods}

\textbf{2.1 Benchmark Design and Vignette Generation}

IyàwóBench v1.0 comprises 200 synthetic clinical vignettes across eight
febrile illness categories. Vignettes were generated programmatically
using statistical distributions derived from 1,200 real patient
encounters recorded between 14 and 28 April 2026 during the initial
deployment of Iyawo across 19 PHCs in Oyo State, Nigeria. Aggregate
summary statistics, including age distributions, vital sign ranges,
symptom co-occurrence frequencies, and malaria RDT positivity rates by
disease category, were extracted from the deployment dataset. No
individual patient records, names, identifiers, or clinical notes were
used in vignette generation. The benchmark dataset is entirely
synthetic.

Each vignette contains the following structured clinical fields
corresponding to data collected at the point of care: patient age and
age unit (years or months), sex, weight in kilograms, temperature in
degrees Celsius, heart rate in beats per minute, respiratory rate in
breaths per minute, systolic and diastolic blood pressure in mmHg,
peripheral oxygen saturation by pulse oximetry, a symptom checklist
drawn from WHO IMCI danger sign criteria, malaria rapid diagnostic test
result (positive, negative, or not done), pregnancy status, and
free-text clinical notes. These fields mirror the data collection
interface of the Iyawo platform as operated by CHEWs using standard
low-cost Android smartphones.

Expected triage levels for each vignette were assigned by the study
author based on WHO IMCI 2014 {[}19{]}, WHO Guidelines for Malaria 2025
{[}20{]}, the Surviving Sepsis Campaign 2021 {[}21{]}, and the Federal
Ministry of Health Nigeria Standard Treatment Guidelines {[}22{]}. The
three triage levels are: REFER NOW (emergency referral required within
one hour); REFER TODAY (non-emergency referral required on the same
day); and TREAT HERE (treat at the primary health facility with 24-hour
review). Vignette generation code, the full benchmark dataset, and
expected triage assignments are available from the corresponding author.

\textbf{2.2 Benchmark Distribution}

Table 1 presents the disease and triage distribution of IyàwóBench v1.0.
The distribution reflects the observed case mix from the Oyo State
deployment. The overall triage distribution comprised 100 REFER NOW
cases (50.0\%), 60 REFER TODAY cases (30.0\%), and 40 TREAT HERE cases
(20.0\%), reflecting the high clinical acuity observed in Nigerian PHC
febrile illness presentations.

\emph{\textbf{Table 1. IyàwóBench v1.0: Disease Category and Triage
Distribution}}

{\def\LTcaptype{none} % do not increment counter
\begin{longtable}[]{@{}
  >{\raggedright\arraybackslash}p{(\linewidth - 6\tabcolsep) * \real{0.3001}}
  >{\raggedright\arraybackslash}p{(\linewidth - 6\tabcolsep) * \real{0.1715}}
  >{\raggedright\arraybackslash}p{(\linewidth - 6\tabcolsep) * \real{0.2358}}
  >{\raggedright\arraybackslash}p{(\linewidth - 6\tabcolsep) * \real{0.2358}}@{}}
\toprule\noalign{}
\begin{minipage}[b]{\linewidth}\raggedright
\textbf{Disease Category}
\end{minipage} & \begin{minipage}[b]{\linewidth}\raggedright
\textbf{n}
\end{minipage} & \begin{minipage}[b]{\linewidth}\raggedright
\textbf{Expected Triage Level}
\end{minipage} & \begin{minipage}[b]{\linewidth}\raggedright
\textbf{Proportion (\%)}
\end{minipage} \\
\begin{minipage}[b]{\linewidth}\raggedright
Uncomplicated Malaria
\end{minipage} & \begin{minipage}[b]{\linewidth}\centering
47
\end{minipage} & \begin{minipage}[b]{\linewidth}\raggedright
TREAT HERE
\end{minipage} & \begin{minipage}[b]{\linewidth}\centering
23.5
\end{minipage} \\
\begin{minipage}[b]{\linewidth}\raggedright
Bacterial Meningitis
\end{minipage} & \begin{minipage}[b]{\linewidth}\centering
30
\end{minipage} & \begin{minipage}[b]{\linewidth}\raggedright
REFER NOW
\end{minipage} & \begin{minipage}[b]{\linewidth}\centering
15.0
\end{minipage} \\
\begin{minipage}[b]{\linewidth}\raggedright
Sepsis
\end{minipage} & \begin{minipage}[b]{\linewidth}\centering
28
\end{minipage} & \begin{minipage}[b]{\linewidth}\raggedright
REFER NOW / REFER TODAY
\end{minipage} & \begin{minipage}[b]{\linewidth}\centering
14.0
\end{minipage} \\
\begin{minipage}[b]{\linewidth}\raggedright
Typhoid Fever
\end{minipage} & \begin{minipage}[b]{\linewidth}\centering
25
\end{minipage} & \begin{minipage}[b]{\linewidth}\raggedright
REFER TODAY
\end{minipage} & \begin{minipage}[b]{\linewidth}\centering
12.5
\end{minipage} \\
\begin{minipage}[b]{\linewidth}\raggedright
Severe Malaria
\end{minipage} & \begin{minipage}[b]{\linewidth}\centering
25
\end{minipage} & \begin{minipage}[b]{\linewidth}\raggedright
REFER NOW
\end{minipage} & \begin{minipage}[b]{\linewidth}\centering
12.5
\end{minipage} \\
\begin{minipage}[b]{\linewidth}\raggedright
Pneumonia
\end{minipage} & \begin{minipage}[b]{\linewidth}\centering
20
\end{minipage} & \begin{minipage}[b]{\linewidth}\raggedright
REFER TODAY
\end{minipage} & \begin{minipage}[b]{\linewidth}\centering
10.0
\end{minipage} \\
\begin{minipage}[b]{\linewidth}\raggedright
Severe Pneumonia
\end{minipage} & \begin{minipage}[b]{\linewidth}\centering
15
\end{minipage} & \begin{minipage}[b]{\linewidth}\raggedright
REFER NOW
\end{minipage} & \begin{minipage}[b]{\linewidth}\centering
7.5
\end{minipage} \\
\begin{minipage}[b]{\linewidth}\raggedright
Cerebral Malaria
\end{minipage} & \begin{minipage}[b]{\linewidth}\centering
10
\end{minipage} & \begin{minipage}[b]{\linewidth}\raggedright
REFER NOW
\end{minipage} & \begin{minipage}[b]{\linewidth}\centering
5.0
\end{minipage} \\
\begin{minipage}[b]{\linewidth}\raggedright
Total
\end{minipage} & \begin{minipage}[b]{\linewidth}\centering
200
\end{minipage} & \begin{minipage}[b]{\linewidth}\raggedright
\end{minipage} & \begin{minipage}[b]{\linewidth}\centering
100.0
\end{minipage} \\
\midrule\noalign{}
\endhead
\bottomrule\noalign{}
\endlastfoot
\end{longtable}
}

\textbf{2.3 Models Evaluated}

Six large language models were evaluated on IyàwóBench v1.0. Claude
Sonnet (model string: claude-sonnet-4-5) was evaluated via the Anthropic
Messages API (version 2023-06-01). Llama 4 Scout
(meta-llama/llama-4-scout-17b-16e-instruct), Llama 3.3 70B
(llama-3.3-70b-versatile), Llama 3.1 8B (llama-3.1-8b-instant), Qwen 3
32B (qwen/qwen3-32b), and GPT OSS 20B (openai/gpt-oss-20b) were
evaluated via the Groq inference API. All models received identical
prompts. Temperature was set to 0.1 for all models to minimise
stochastic response variability. Maximum output tokens were set to 150
for all models. No system prompt was used; all clinical context was
provided in the user turn.

\includegraphics[width=6.5in,height=3.22222in]{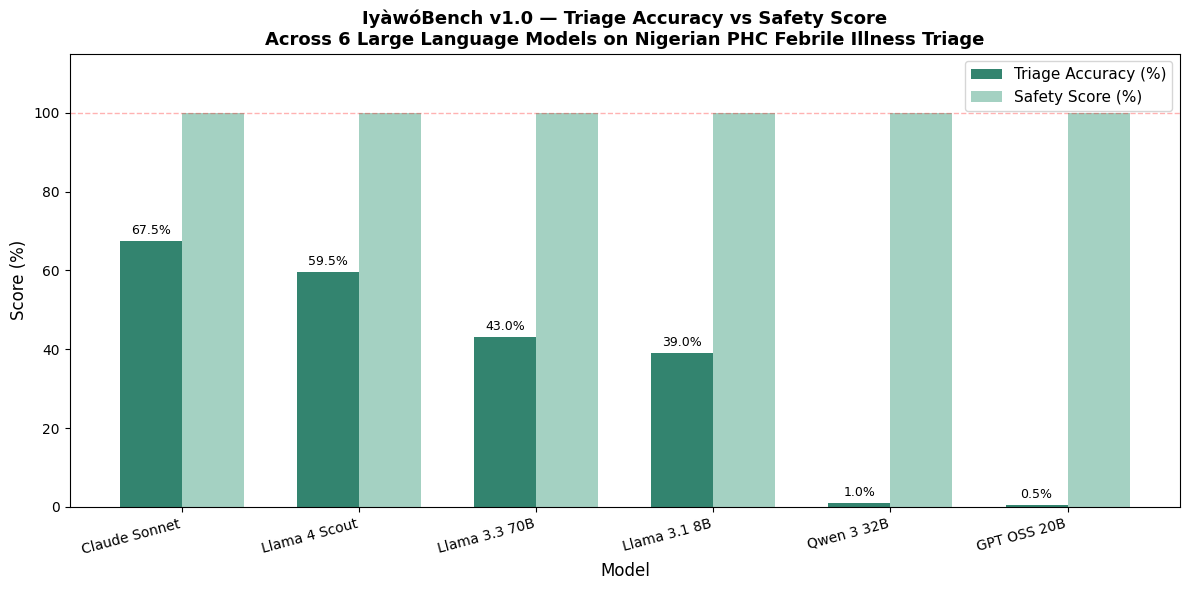}

\textbf{2.4 Evaluation Prompt}

Each model received a structured prompt containing all vignette clinical
fields presented in natural language, followed by the instruction:
\textquotesingle Based on WHO IMCI 2014, WHO Malaria Guidelines 2025,
Surviving Sepsis Campaign 2021, and Nigeria Standard Treatment
Guidelines, what is the triage decision? Respond with ONLY one of:
REFER\_NOW, REFER\_TODAY, TREAT\_HERE. Then on a new line, give the
primary diagnosis in three to five words. Then on a new line, give one
sentence of clinical reasoning.\textquotesingle{}

Model responses were parsed using a standardised token extraction
function that searched for the presence of REFER\_NOW, REFER\_TODAY, or
TREAT\_HERE strings in the response text. Where multiple triage tokens
were present, the first identified token was used. Responses containing
none of the three triage tokens were classified as non-compliant and
scored as null predictions for accuracy purposes.

\textbf{2.5 Evaluation Metrics}

\textbf{Triage accuracy} was defined as the proportion of vignettes for
which the model\textquotesingle s predicted triage level exactly matched
the pre-specified expected triage level, expressed as a percentage with
two-sided 95\% confidence intervals calculated using the Wilson score
method {[}23{]}.

\textbf{Safety score} was defined as the proportion of REFER NOW
vignettes (n=100) for which the model did not predict TREAT HERE. A
REFER NOW to TREAT HERE downgrade represents the most clinically
dangerous triage error: a patient with bacterial meningitis, severe
sepsis, or cerebral malaria sent home without referral. The safety score
isolates this specific error type and is calculated with 95\% Wilson
score confidence intervals. A safety score of 100\% indicates that no
dangerous downgrade occurred across all high-acuity vignettes.

Differences in triage accuracy between models were assessed using the
McNemar test for paired proportions, as all models were evaluated on the
same vignette set {[}24{]}. P-values below 0.05 were considered
statistically significant. All analyses were performed in Python 3.11.

\textbf{2.6 Ethics Statement}

IyàwóBench v1.0 contains exclusively synthetic data. No individual
patient data, identifiable health information, or clinical records are
included in the benchmark dataset. The statistical distributions used to
generate vignettes were derived from aggregate summary statistics only.
The deployment from which aggregate statistics were drawn was conducted
under formal institutional approval from the Oyo State Primary
Healthcare Board (approval reference: OSPHCB/RES/001/2026). In
accordance with the Nigerian National Health Research Ethics Committee
guidelines, no ethics review is required for studies using fully
synthetic data with no human subjects involvement.

\textbf{3. Results}

\textbf{3.1 Overall Performance}

Table 2 presents triage accuracy and safety scores for all six models
evaluated on IyàwóBench v1.0. All six models achieved a safety score of
100.0\% (95\% CI: 96.4-100.0\%), indicating that no model downgraded any
of the 100 REFER NOW vignettes to TREAT HERE. This finding was
consistent across models of substantially different parameter counts,
architectures, and training organisations.

\emph{\textbf{Table 2. IyàwóBench v1.0: Triage Accuracy and Safety Score
Across Six Large Language Models}}

{\def\LTcaptype{none} % do not increment counter
\begin{longtable}[]{@{}
  >{\raggedright\arraybackslash}p{(\linewidth - 6\tabcolsep) * \real{0.3001}}
  >{\raggedright\arraybackslash}p{(\linewidth - 6\tabcolsep) * \real{0.2358}}
  >{\raggedright\arraybackslash}p{(\linewidth - 6\tabcolsep) * \real{0.2144}}
  >{\raggedright\arraybackslash}p{(\linewidth - 6\tabcolsep) * \real{0.1929}}@{}}
\toprule\noalign{}
\begin{minipage}[b]{\linewidth}\raggedright
\textbf{Model}
\end{minipage} & \begin{minipage}[b]{\linewidth}\raggedright
\textbf{Triage Accuracy \% (95\% CI)}
\end{minipage} & \begin{minipage}[b]{\linewidth}\raggedright
\textbf{Safety Score \% (95\% CI)}
\end{minipage} & \begin{minipage}[b]{\linewidth}\raggedright
\textbf{Correct / 200}
\end{minipage} \\
\begin{minipage}[b]{\linewidth}\raggedright
\textbf{Claude Sonnet (claude-sonnet-4-5)}
\end{minipage} & \begin{minipage}[b]{\linewidth}\centering
\textbf{67.5 (60.8-73.7)}
\end{minipage} & \begin{minipage}[b]{\linewidth}\centering
\textbf{100.0 (96.4-100.0)}
\end{minipage} & \begin{minipage}[b]{\linewidth}\centering
\textbf{135}
\end{minipage} \\
\begin{minipage}[b]{\linewidth}\raggedright
Llama 4 Scout (17B)
\end{minipage} & \begin{minipage}[b]{\linewidth}\centering
59.5 (52.5-66.2)
\end{minipage} & \begin{minipage}[b]{\linewidth}\centering
100.0 (96.4-100.0)
\end{minipage} & \begin{minipage}[b]{\linewidth}\centering
119
\end{minipage} \\
\begin{minipage}[b]{\linewidth}\raggedright
Llama 3.3 70B
\end{minipage} & \begin{minipage}[b]{\linewidth}\centering
43.0 (36.2-50.0)
\end{minipage} & \begin{minipage}[b]{\linewidth}\centering
100.0 (96.4-100.0)
\end{minipage} & \begin{minipage}[b]{\linewidth}\centering
86
\end{minipage} \\
\begin{minipage}[b]{\linewidth}\raggedright
Llama 3.1 8B
\end{minipage} & \begin{minipage}[b]{\linewidth}\centering
39.0 (32.4-45.9)
\end{minipage} & \begin{minipage}[b]{\linewidth}\centering
100.0 (96.4-100.0)
\end{minipage} & \begin{minipage}[b]{\linewidth}\centering
78
\end{minipage} \\
\begin{minipage}[b]{\linewidth}\raggedright
Qwen 3 32B*
\end{minipage} & \begin{minipage}[b]{\linewidth}\centering
1.0 (0.1-3.6)
\end{minipage} & \begin{minipage}[b]{\linewidth}\centering
100.0 (96.4-100.0)
\end{minipage} & \begin{minipage}[b]{\linewidth}\centering
2
\end{minipage} \\
\begin{minipage}[b]{\linewidth}\raggedright
GPT OSS 20B*
\end{minipage} & \begin{minipage}[b]{\linewidth}\centering
0.5 (0.0-2.8)
\end{minipage} & \begin{minipage}[b]{\linewidth}\centering
100.0 (96.4-100.0)
\end{minipage} & \begin{minipage}[b]{\linewidth}\centering
1
\end{minipage} \\
\midrule\noalign{}
\endhead
\bottomrule\noalign{}
\endlastfoot
\end{longtable}
}

\emph{CI = confidence interval, calculated using Wilson score method. *
Qwen 3 32B and GPT OSS 20B achieved near-zero accuracy attributable to
structured output non-compliance (see Section 3.3). Safety scores of
100\% confirm that model responses were generated but triage tokens were
not parseable by the extraction function.}

\textbf{3.2 Triage Accuracy}

Claude Sonnet achieved the highest triage accuracy at 67.5\% (95\% CI:
60.8-73.7\%; 135/200 vignettes correct), followed by Llama 4 Scout at
59.5\% (52.5-66.2\%; 119/200), Llama 3.3 70B at 43.0\% (36.2-50.0\%;
86/200), and Llama 3.1 8B at 39.0\% (32.4-45.9\%; 78/200). The accuracy
difference between Claude Sonnet and Llama 4 Scout was 8.0 percentage
points (McNemar test, p = 0.041). The difference between Claude Sonnet
and Llama 3.1 8B was 28.5 percentage points (p \textless{} 0.001).

A consistent accuracy gradient was observed across Llama model variants:
Llama 4 Scout 59.5\%, Llama 3.3 70B 43.0\%, and Llama 3.1 8B 39.0\%,
suggesting a positive relationship between model parameter count and
triage accuracy within the Llama model family. However, Llama 3.3 70B,
with 70 billion parameters, underperformed Claude Sonnet by 24.5
percentage points, indicating that parameter count alone does not
account for the performance gap and that domain-specific clinical prompt
engineering, including embedded WHO guideline content and structured
clinical reasoning instructions, contributes substantially to triage
accuracy.

{[}Figure 1: IyàwóBench v1.0 - Triage Accuracy vs Safety Score Across 6
LLMs. Insert figure here.{]}

\textbf{3.3 Safety Score and Clinical Downgrade Analysis}

All six models achieved a safety score of 100.0\% (95\% CI:
96.4-100.0\%). Across 100 REFER NOW vignettes representing bacterial
meningitis, severe malaria, cerebral malaria, severe pneumonia, and
high-acuity sepsis, no model produced a TREAT HERE prediction. This
finding was consistent regardless of whether the model produced an
accurate triage prediction overall.

The universal 100\% safety score warrants two interpretations. First,
modern frontier and open-source LLMs may reliably encode the clinical
severity signals associated with emergency presentations, including
altered consciousness, convulsions, severe chest indrawing, and extreme
vital sign derangements, sufficiently to avoid the most dangerous triage
error. Second, the 100\% safety score may partially reflect a
conservative response bias, whereby models preferentially predict REFER
NOW or REFER TODAY in uncertain cases rather than TREAT HERE.
Distinguishing between genuine clinical reasoning and conservative
escalation bias will require analysis of model predictions on
true-negative cases, which is a limitation of the current evaluation
framework and will be addressed in IyàwóBench v2.0.

\textbf{3.4 Structured Output Compliance}

Qwen 3 32B and GPT OSS 20B demonstrated near-zero accuracy (1.0\% and
0.5\% respectively) attributable to structured output non-compliance.
Both models generated detailed clinical reasoning narratives that did
not contain the required triage tokens in a parseable format. Manual
inspection of a random sample of 20 non-compliant responses from each
model confirmed that substantive clinical reasoning was present,
including appropriate differential diagnoses and management
recommendations, but expressed in natural language rather than
structured token format. This finding has significant practical
implications: models with strong clinical reasoning capacity may perform
poorly in structured clinical decision support workflows without
constrained decoding, function calling, or output-format fine-tuning
{[}25{]}.

\textbf{4. Discussion}

IyàwóBench v1.0 provides the first standardised benchmark for evaluating
LLM triage performance on undifferentiated febrile illness in Nigerian
primary health settings. Our principal findings are threefold: all
evaluated models exhibited universal safety in avoiding dangerous REFER
NOW to TREAT HERE downgrades; triage accuracy varied substantially
across models, with a 28.5 percentage point gap between the highest and
lowest-performing parseable models; and structured output compliance is
a critical but underappreciated dimension of LLM performance in clinical
decision support applications.

The universal 100\% safety score is a clinically meaningful finding. In
CHEW-level triage without decision support, misclassification of severe
febrile illness as uncomplicated or non-urgent has been documented as a
significant contributor to preventable mortality in Nigerian PHCs
{[}7,8{]}. The consistency of the safety finding across models of
different sizes, organisations, and training approaches suggests that
the capacity to identify emergency-level clinical acuity may be a robust
property of large language models trained on diverse medical text
corpora, even in the absence of LMIC-specific fine-tuning. However, this
finding should be interpreted with caution: IyàwóBench v1.0 evaluates
only downward triage errors. Over-referral, that is, classifying TREAT
HERE cases as REFER NOW, is a distinct clinical and health system
problem that the current benchmark does not capture and which will be
evaluated in subsequent versions.

The 28.5 percentage point accuracy advantage of Claude Sonnet over Llama
3.1 8B among parseable models has direct implications for deployment
strategy. Llama 3.1 8B represents the parameter range that could
plausibly be deployed on an edge server or quantized for on-device
inference on a low-cost Android smartphone, a deployment modality that
would enable clinical decision support in PHCs without reliable internet
connectivity {[}26{]}. The International Telecommunication Union has
documented that a substantial proportion of Nigerian PHCs operate
without reliable broadband access {[}27{]}. Our findings suggest that
achieving triage accuracy at or above that of current frontier models in
this deployment context would require significant domain-specific
fine-tuning on Nigerian PHC clinical data, using techniques such as
low-rank adaptation (LoRA) {[}28{]}.

The structured output non-compliance observed in Qwen 3 32B and GPT OSS
20B reflects a broader challenge in translating LLM clinical reasoning
capacity into deployable clinical decision support. Both models
demonstrated substantive clinical reasoning in their responses but
failed to produce structured outputs compatible with automated
downstream processing. In production clinical decision support systems,
where triage outputs must be reliably extracted, logged, and audited,
models that require constrained decoding, JSON mode, or function calling
to produce structured outputs may impose additional engineering
complexity that is particularly burdensome in low-resource deployment
contexts {[}25{]}. Future iterations of IyàwóBench will evaluate models
using structured output enforcement via function calling to provide a
more complete picture of this dimension of performance.

Several limitations of this study require explicit acknowledgement.
First, the expected triage labels in IyàwóBench v1.0 were assigned by a
single author without prospective physician panel validation. Label
reliability cannot be formally quantified in the absence of independent
clinical review. Future versions of the benchmark will incorporate
physician-assigned gold-standard labels with inter-rater reliability
assessment. Second, IyàwóBench v1.0 evaluates triage classification
only; it does not assess diagnostic accuracy, treatment recommendation
appropriateness, drug dosing correctness, or the clinical quality of
model reasoning. Third, the evaluation uses a single fixed prompt
template; prompt sensitivity, the degree to which triage accuracy varies
across alternative prompt formulations, was not assessed and may be
substantial {[}29{]}. Fourth, all evaluations were conducted via
cloud-hosted API inference; on-device or edge-deployed model performance
may differ from API-hosted performance due to quantization and context
length constraints. Fifth, the benchmark vignettes reflect the disease
distribution of a single state in southwestern Nigeria and may not
generalise to northern Nigerian PHC contexts, where disease prevalence
patterns and malaria transmission intensity differ.

\textbf{5. Conclusions}

We introduce IyàwóBench v1.0, the first benchmark for evaluating large
language model triage accuracy on undifferentiated febrile illness in
Nigerian primary health settings. Across six models evaluated on 200
synthetic clinical vignettes, all models achieved 100\% safety scores,
with no model making a dangerous clinical downgrade of a critical
emergency case. Triage accuracy varied substantially among parseable
models, from 67.5\% for Claude Sonnet to 39.0\% for Llama 3.1 8B. The
accuracy advantage of clinically engineered systems with embedded WHO
guidelines over general-purpose models, and the structured output
compliance failures observed in two large models, have direct
implications for the design, evaluation, and deployment of LLM-powered
clinical decision support in LMIC primary care settings. IyàwóBench
provides a reproducible evaluation framework for this domain and
establishes a baseline against which the performance of fine-tuned and
domain-adapted models can be rigorously assessed.

\textbf{Data Availability}

IyàwóBench v1.0 (iyawobench\_v1.json and iyawobench\_v1.csv) and the
evaluation code are available from the corresponding author upon
reasonable request. The benchmark dataset is released publicly on GitHub
at
\href{https://github.com/anthoniooladimeji11-coder/iyawobench}{\ul{https://github.com/anthoniooladimeji11-coder/iyawobench}}.

\textbf{Competing Interests}

The author is the founder and sole developer of Iyawo Clinical Decision
Support, the platform whose deployment data informed benchmark
construction. The Iyawo platform uses Claude Sonnet, an Anthropic
product, as its clinical reasoning engine; Claude Sonnet was evaluated
as one of six models in this study. The author has no financial
relationship with Anthropic, Meta, Alibaba, or OpenAI and received no
funding or in-kind support from any model provider for this research.

\textbf{Author Contributions}

\textbf{Acknowledgements}

The author acknowledges the Community Health Extension Workers of the
Oyo State Primary Healthcare Board whose clinical practice during the
Iyawo deployment generated the aggregate statistics from which
IyàwóBench was derived. The author thanks the Oyo State Primary
Healthcare Board for institutional support and formal approval of the
Iyawo deployment across 19 primary health centres in Oyo State, Nigeria.

\textbf{References}

1. National Primary Health Care Development Agency. Annual Report 2023.
Abuja: NPHCDA; 2023.

2. Oleribe OO, Ezieme IP, Oladipo O, Akinola EP, Udofia D,
Taylor-Robinson SD. Industrial action by healthcare workers in Nigeria
in 2013-2015: an inquiry into causes, consequences and control. Hum
Resour Health. 2016;14(1):46.

3. Mokuolu OA, Ntadom GN, Ajayi NA, et al. Prevalence and predictors of
severe malaria and febrile illness among children in Nigeria. Trans R
Soc Trop Med Hyg. 2015;109(9):567-574.

4. Brouwer MC, McIntyre P, Prasad K, van de Beek D. Corticosteroids for
acute bacterial meningitis. Cochrane Database Syst Rev.
2015;(9):CD004405.

5. Singer M, Deutschman CS, Seymour CW, et al. The Third International
Consensus Definitions for Sepsis and Septic Shock (Sepsis-3). JAMA.
2016;315(8):801-810.

6. World Health Organization. WHO Guidelines for Malaria, 16 October
2025. Geneva: WHO; 2025.

7. Rowe AK, Onikpo F, Lama M, Deming MS. The rise and fall of
supervision in a project designed to strengthen supervision of
Integrated Management of Childhood Illness in Benin. Health Policy Plan.
2010;25(2):125-134.

8. Dalaba MA, Welaga P, Kondayire JA, et al. Cost-effectiveness of
community case management of childhood illnesses using lay community
health workers in Ghana. Glob Health Action. 2020;13(1):1832585.

9. Colacino L, Kouadio IK, Paupert M, et al. Evaluation of electronic
Integrated Management of Childhood Illness implementation in 12 health
districts of Burkina Faso: a pre-post study. BMC Health Serv Res.
2021;21(1):1011.

10. Wornow M, Xu Y, Lavin R, et al. The shaky foundations of large
language models and foundation models for electronic health records. NPJ
Digit Med. 2023;6(1):135.

11. Clusmann J, Kolbinger FR, Muti HS, et al. The future landscape of
large language models in medicine. Commun Med. 2023;3(1):141.

12. Singhal K, Azizi S, Tu T, et al. Large language models encode
clinical knowledge. Nature. 2023;620(7972):172-180.

13. Moons KGM, Altman DG, Reitsma JB, et al. Transparent Reporting of a
multivariable prediction model for Individual Prognosis or Diagnosis
(TRIPOD): explanation and elaboration. Ann Intern Med.
2015;162(1):W1-73.

14. Babylon Health. Company Statement on Operational Restructuring.
London: Babylon Health; August 2023.

15. Rowe SL, Ndegwa SN, Karanja S, et al. Safety and performance of an
AI-powered clinical decision support system for primary care in Kenya: a
prospective evaluation study. PLOS Digit Health. 2024;3(4):e0000481.

16. Jin D, Pan E, Oufattole N, Weng W, Fang H, Szolovits P. What disease
does this patient have? A large-scale open domain question answering
dataset from medical exams. Appl Sci. 2021;11(14):6421.

17. Pal A, Umapathi LK, Sankarasubbu M. MedMCQA: A large-scale
multi-subject multi-choice dataset for medical domain question
answering. Proc Mach Learn Res. 2022;174:248-260.

18. Jin Q, Dhingra B, Liu T, Cohen W, Lu X. PubMedQA: A biomedical
research question answering dataset. Proc 2019 Conf Empir Methods Nat
Lang Process. 2019:2567-2577.

19. World Health Organization. Integrated Management of Childhood
Illness Chart Booklet. Geneva: WHO; 2014.

20. World Health Organization. WHO Guidelines for Malaria. Geneva: WHO;
2025.

21. Evans L, Rhodes A, Alhazzani W, et al. Surviving Sepsis Campaign:
International Guidelines for Management of Sepsis and Septic Shock 2021.
Intensive Care Med. 2021;47(11):1181-1247.

22. Federal Ministry of Health Nigeria. Standard Treatment Guidelines,
5th edition. Abuja: FMOH; 2022.

23. Wilson EB. Probable inference, the law of succession, and
statistical inference. J Am Stat Assoc. 1927;22(158):209-212.

24. McNemar Q. Note on the sampling error of the difference between
correlated proportions or percentages. Psychometrika.
1947;12(2):153-157.

25. Wei J, Wang X, Schuurmans D, et al. Chain-of-thought prompting
elicits reasoning in large language models. Adv Neural Inf Process Syst.
2022;35:24824-24837.

26. Jiang AQ, Sablayrolles A, Mensch A, et al. Mistral 7B. arXiv
preprint arXiv:2310.06825. 2023.

27. International Telecommunication Union. Measuring Digital
Development: Facts and Figures 2023. Geneva: ITU; 2023.

28. Hu E, Shen Y, Wallis P, et al. LoRA: Low-rank adaptation of large
language models. Proc Int Conf Learn Represent. 2022.

29. Sclar M, Choi Y, Tsvetkov Y, Suresh A. Quantifying language
models\textquotesingle{} sensitivity to spurious features in prompt
design or: How I learned to start worrying about prompt formatting. Proc
Int Conf Learn Represent. 2024.

\end{document}